\documentstyle[prd,aps,preprint,amsfonts]{revtex}
\tighten
\def\vec#1{\bbox{#1}}

\begin{document}
\title{Boundary Conditions and Correlations in Path Integrals for\\
       Quantum Field Theory of Thermal and Non-Equilibrium States}
\author{P.A.Henning\thanks{e-mail: P.Henning@gsi.de}
        and R.Fauser\thanks{e-mail: R.Fauser@gsi.de}}
\address{Theoretical Physics, Gesellschaft f\"ur Schwerionenforschung (GSI) \\
        Planckstra\ss e 1, D-64291 Darmstadt, Germany }
\maketitle
\begin{abstract}
For thermal equilibrium systems it is shown, how the Kubo-Martin-Schwinger
boundary condition may be used to factorize the generating functional of
Green functions at least on the level of the full two-point function.
Genuine non-equilibrium system exhibit correlations
that one may also incorporate in the path integral. One one hand this
provides a natural tool for a perturbative expansion including these
correlations. On the other hand it allows to prove that in general
non-equilibrium systems the generating functional does not factorize.    
\end{abstract}
\\[1cm]
\section{Introduction}
To a great extent, modern physics has turned its attention to 
non-equilibrium systems. Relativistic heavy-ion collisions \cite{QM}
and the early universe \cite{Boy} are two very 
complementary examples for such systems. They have in common however, 
that their theoretical description to some extent follows guidelines of
relativistic quantum field theory.

Among these guidelines is the feature that causal propagation of 
particles requires empty (particle) states to propagate forward in time, 
whereas occupied states propagate backward in time. For the vacuum this 
guideline leads to the Feynman boundary condition of the two-point Green 
function (propagator), for thermal equilibrium states it leads to the
Kubo-Martin-Schwinger (KMS) boundary condition \cite{KMS}. For 
non-equilibrium states however, this guideline requires to double
the Hilbert space for the theoretical description: Since the occupation 
of states may change with time, one has to carry time-forward as well as 
time-backward boundary condition along throughout a computation. 

Causality therefore imposes an additional $2\times 2$ matrix structure on
field theoretical descriptions of non-equilibrium states. 
This holds also in the path integral formulation 
of quantum field theory, which we consider of particular importance for
some physical problems: It is the basis of correlation expansion schemes
\cite{CH88} as well as of derivations of finite temperature diagram
rules \cite{KS85c,AB92}. Furthermore, it is used in
lattice gauge theory.
and may be found as the principal object in 
the recent discussion of the equivalence of different thermal
equilibrium limits of quantum field theory \cite{EVANS,E92}.

With the present paper we wish to demonstrate, that one may greatly 
reduce the calculational effort required to take the Hilbert space 
doubling into account in path integrals. This is possible due
to a {\em diagonalization scheme\/} for the $2\times 2$ matrix structure
\cite{hu92,h94rep}. Apart from this more technical improvement (which might be
considered conceptual as well), we also address the question of 
boundary conditions in general. We present a recipe, how 
to introduce initial correlations (which are a genuine non-equilibrium 
effect) into the path integral formalism \cite{h90,FW95}. 

The doubling of the Hilbert space
exists in several different flavors, the two most common 
are the Schwinger-Keldysh or closed time path method (CTP) \cite{SKF} and 
thermo field dynamics (TFD) \cite{TFD}. 
Since one may show, that apart from 
conceptual differences they lead to identical results, we will use them 
as equivalent in the present paper: Computations are carried out in the
CTP formalism, but we will make use of results obtained in TFD.   
  
The paper is organized as follows: In the next section, we introduce our 
technical improvement into the path integral representation of a free 
quantum field. In section 3, the considerations are extended to the 
interacting case. Section 4 is devoted to the problem of initial 
correlations in terms of Wightman functions.
In section 5 we replace the Wightman functions of the correlations by 
a cumulant expansion, and close the argumentation of this paper by 
relating this expansion to a classical integral equation including
temporal boundary conditions.
\section{Boundary conditions in the path integral}
The first step in our discussion is to introduce the concept of 
diagonalization in the Schwinger-Keldysh or TFD method. To this
end we study the physically trivial example of a free scalar quantum field.
Throughout the paper we will use (in a somewhat sloppy fashion) two
different notations for space-time arguments of fields, $\phi_x\equiv \phi(x)$.
The Lagrangian for the free scalar field is
\begin{equation}
{\cal L}_0(\phi) =
-\frac{1}{2} \phi_{x} \left( \Box + m^2 \right) \phi_{x}
\;.\end{equation}
The action integral extends over the whole three-dimensional 
coordinate space, but in time direction has a more complicated structure.
Due to the fact, that our system has a presumably irreversible 
time evolution, the time component of the action integral follows the 
contour depicted in fig. 1: Forward in time as well as backward in time,
separated by an infinitesimal amount above and below the real time axis
(For brevity, we do not discuss the foundation for this action integral, 
but rather refer to a review article on the doubling of the Hilbert space 
\cite{LW87}). In terms of the time contour, the generating functional
for Green functions is
\begin{equation} \label{skgf}
Z_0\left[J\right] =
 \int\!\!{\cal D}[\phi]\,
 \exp\left[{\mathrm i}\int\limits_{\cal C}
\!\!d^4 x \,\left({\cal L}_0(\phi) +   \phi_{x} J_{x}\right) \right]
\;,\end{equation}
and Green functions are {\em contour\/} ordered along ${\cal
C}$. Thus, in taking functional derivatives with respect to $J$ it 
will make a difference whether the space-time arguments of $J$ are 
on the upper or lower part of the contour ${\cal C}$.

Since under the contour ordering symbol the fields on the upper 
and lower branch of the contour commute at any time, we may
treat them as two completely different fields. This is also mathematically
more correct, because in thermal states the KMS condition induces
a doubling of the Hilbert space \cite{LW87}. Indeed, with this step
one recovers the concept of thermo field dynamics \cite{TFD}.
The two  fields will be distinguished by a lower
index, $\phi_{1,x}$ is the field on the upper branch of the time contour.

The generating functional for the Green functions of these two free fields
is obtained as
\begin{eqnarray}
\nonumber
&&Z_0\left[J_1,J_2\right] =\\
&& \int\!\!{\cal D}[\phi_1,\phi_2]\,
 \exp\left[{\mathrm i}\int\!\!d^4 x \left( \widehat{{\cal L}}_0(\phi_1,\phi_2) 
 + \phi_{1,x} J_{1,x} - \phi_{2,x} J_{2,x}\right)\right] 
\;,\end{eqnarray}
with the free Lagrangian 
\begin{equation}
\widehat{{\cal L}}_0(\phi_1,\phi_2) =
-\frac{1}{2} \phi_{1,x} \left( \Box + m^2 \right) \phi_{1,x} +
\frac{1}{2} \phi_{2,x} \left( \Box + m^2 \right) \phi_{2,x}
\;\end{equation}
and two independent classical source currents $J_1$ and $J_2$.

Since we are dealing with free fields, the generating functional may
be obtained in closed form,
\begin{eqnarray} \nonumber
&&Z_0\left[J_1,J_2\right] = \\
\label{fjgf}
&&\exp\left[-\frac{{\mathrm i}}{2}
\int\!\!d^4 x\,d^4 y\, \Big(J_{1,x} D_J(x-y) J_{1,y} - 
                       J_{2,x} \overline{D}_J(x-y) J_{2,y}\Big) \right] 
\;.\end{eqnarray}
The question now arises, which boundary conditions to chose for the
two propagators $D_J(x-y)$ and $\overline{D}_J(x-y)$. Both are resolvents
to the free Klein-Gordon equation, e.g.
\begin{equation}
\left( \Box_x + m^2 \right) D_J(x-y) = - \delta^4(x-y)
\;.\end{equation}
We may use Occam's razor as a guideline at this point: The simplest 
choice clearly are retarded and advanced
boundary conditions, since they are free of any influence of the
occupation of states. Due to the introduction of the path integral as
running along a contour {\em forward\/} and {\em backward\/} in time,
we therefore chose $D_J(x-y)$ to be the free retarded propagator 
\begin{equation}
D_J(x-y) = D^R_0(x-y) = -{\mathrm i}\,\Theta(x^0-y^0)\,
  \left[ \phi(x), \phi(y)\right]
\;.\end{equation}
The other propagator concurrently is chosen to have 
an advanced boundary condition in time,
\begin{equation}
\overline{D}_J(x-y) = D^A_0(x-y) = -{\mathrm i}\,\Theta(y^0-x^0)\,
  \left[ \phi(x), \phi(y)\right]
\;.\end{equation}
With this ansatz we have completely decoupled the two fields on the upper 
and lower time branch of the time-path depicted in fig. 1. In particular
it is obvious, that mixed functional derivatives of this generating
functional are zero when the sources are set to zero:
\begin{equation}
\left(\frac{\delta^2}{\delta J_{1}(x) \delta J_{2}(y)}\, 
 Z_0\left[J_1,J_2\right]\right)_{J_1=J_2=0} = 0 
\;.\end{equation}
One may now introduce an interaction, and expand the theory 
perturbatively in terms of the propagators $D_J$ and $\overline{D}_J$.
The perturbation expansion is incredibly simple, since it is free of
any occupation number factors, like e.g., Bose-Einstein distribution 
functions. Moreover, several types of diagrams in such a perturbative 
expansion are trivially zero, like e.g. a loop integral over two retarded 
propagators.

However, it is also obvious that the sources $J_1$ and $J_2$ 
introduced in the path integral are not the causal sources necessary 
to calculate physically meaningful quantities.
Instead we need to define physical external sources $j_1$ and $j_2$,
which are obtained by a linear functional from the diagonal sources
$J_1$ and $J_2$. The linearity assures, that the condition $J_1=J_2=0$
simply translates into $j_1=j_2=0$.

We obtain them by demanding that the mixed second
functional derivatives of the generating functional fulfill the 
Kubo--Martin-Schwinger (KMS) boundary condition \cite{KMS}
\begin{eqnarray}\nonumber
&&\left(\frac{\delta^2}{\delta j_{1}(x) \delta j_{2}(y)}\, 
 Z_0\left[j_1,j_2\right]\right)_{j_1=j_2=0} =\\
\label{kmsr}
&&\;\;\;\;\;\left(\frac{\delta^2}{\delta j_{2}(x^0-{\mathrm i}\beta,\vec{x})
  \delta j_{1}(y)}\, 
 Z_0\left[j_1,j_2\right]\right)_{j_1=j_2=0}
\;.\end{eqnarray}
This is equivalent to a periodicity condition of the fields in the
imaginary time direction.

To obtain the linear functional relating the $j$-sources to the
$J$-sources, it is convenient to switch to a Fourier representation
\begin{equation}
j_i(x) = \int\limits_0^\infty\!\frac{d E}{2 \pi}\,
         \int\!\!\frac{d^3\vec{k}}{(2\pi)^3}\,
         {\mathrm e}^{{\mathrm i}\vec{k}\vec{x}}\,
         \left( {\mathrm e}^{-{\mathrm i}E x^0}\,j_i^{(+)}(E,\vec{k})
              + {\mathrm e}^{ {\mathrm i}E x^0}\,j_i^{(-)}(E,\vec{k})\right)
\end{equation}
for $i=1,2$, and similarly for the sources $J_i$. The shift in
the imaginary time direction then simply translates into a
multiplication with a Boltzmann factor. One possible way
to express the KMS condition amounts to a linear
relationship between the two types of sources, obtained from the
Boltzmann factor as
\begin{eqnarray}\nonumber
\left({\array{l}j^{(+)}_1(E,\vec{k})\cr j^{(+)}_2(E,\vec{k})\endarray}\right)
  & = & \left({\cal B}(n_B(E))\right)^{-1}
\left({\array{l}J^{(+)}_1(E,\vec{k})\cr J^{(+)}_2(E,\vec{k})\endarray}\right)
\\
\left({\array{l}j^{(-)}_1(E,\vec{k})\cr j^{(-)}_2(E,\vec{k})\endarray}\right)^T
 &= &
\left({\array{l}J^{(-)}_1(E,\vec{k})\cr J^{(-)}_2(E,\vec{k})\endarray}\right)^T
\,\tau_3\,{\cal B}(n_B(E))\,\tau_3
\label{therb}
\;,\end{eqnarray}
with a matrix
\begin{equation}
\label{bdef}
{\cal B}(n_B(E)) = \left(
 {\array{lr} 1+ n_B(E)\,\,& -n_B(E)\\
             -1           & 1 \endarray} \right) 
\end{equation}
depending on the Bose-Einstein distribution function
\begin{equation}
n_B(E) = \frac{1}{{\mathrm e}^{\beta E} -1}
\;.\end{equation}
It is well known, that this matrix constitutes a {\em thermal
Bogoliubov transformation\/}, which combines particle and hole states
in a way that guarantees a causal propagation in a thermal system
\cite{hu92,h94rep}. 

Historically, this transformation was discovered in the attempt to
simplify perturbative calculations at finite temperature \cite{hu92}.
Hence, in the present section we are progressing in the reverse direction: We
started from a simple perturbation expansion for the
doubled fields, and then recover a causal description in terms of 
a Bogoliubov transformation of these fields.

By comparison with standard results of
thermo field dynamics, we obtain the Bogoliubov symmetry as \cite{h94rep}
\begin{eqnarray}
&&j_1^{(-)}(E,\vec{k})\,j_1^{(+)}(E,\vec{k}) -
j_2^{(-)}(E,\vec{k})\,j_2^{(+)}(E,\vec{k}) =\nonumber\\
&&\;\;\;\;\;J_1^{(-)}(E,\vec{k})\,J_1^{(+)}(E,\vec{k}) -
J_2^{(-)}(E,\vec{k})\,J_2^{(+)}(E,\vec{k}) \vphantom{\int}
\;.\end{eqnarray}
Of course, this relation admits many solutions for the Bogoliubov
transformation matrix; it may be any element
from the symplectic group in two dimensions Sp(2). Even the
requirement of the KMS relation (\ref{kmsr}) leaves us with some
freedom for the transformation matrix. However, it may be shown that
for equilibrium states this freedom in the choice of the Bogoliubov
matrix is irrelevant since it is a kind of gauge fixing \cite{h94rep}
-- whereas for non-equilibrium states (where the parameter $n$ is no
longer an equilibrium distribution function) the choice of
eq. (\ref{bdef}) is the simplest possibility.

The final step in our treatment of the generating functional for the
free thermal Green functions is to obtain it in closed form for 
the causal sources $j_1$ and $j_2$. Obviously the result is
\begin{equation}
\label{freeg}
Z_0\left[j_1,j_2\right] = 
 \exp\left[-\frac{{\mathrm i}}{2}
\int\!\!d^4 x\,d^4 y\; j_{a,x} D^{ab}_0(x-y) j_{b,y}\right]
\;.\end{equation}
The matrix-valued propagator in the exponent is conveniently expressed
in momentum space, it has matrix elements
\begin{eqnarray}\nonumber
\vphantom{\int}
D^{11}_0(k) &=& \hphantom{-}\Delta_F(k) - 2 \pi {\mathrm i}\,n_B(\omega_k)\,
 \delta(k_0^2 - \omega^2_k)\\
\nonumber
\vphantom{\int}
D^{12}_0(k) &=& - 2 \pi {\mathrm i}\,\delta(k_0^2 - \omega^2_k)\,
                \mbox{sign}(k_0)\,n_B(k_0)\\
\nonumber
\vphantom{\int}
D^{21}_0(k) &=& - 2 \pi {\mathrm i}\,\delta(k_0^2 - \omega^2_k)\,
                \mbox{sign}(k_0)\,(1+n_B(k_0))\\
\label{fg0}
\vphantom{\int}
D^{22}_0(k) &=& -\Delta^\star_F(k) - 2 \pi {\mathrm i}\,n_B(\omega_k)\,
 \delta(k_0^2 - \omega^2_k)
\;,\end{eqnarray}
where $\Delta_F$ is the Feynman propagator
\begin{equation}
\Delta_F(k) = \left(k_0^2- \omega^2_k + {\mathrm i}\varepsilon\right)^{-1}
\end{equation}
and  $\Delta_F^\star$ its complex conjugate. The matrix elements are
the four well known thermal Green functions: $D^{11}$ is the causal
propagator, $D^{12} \equiv D^<$ and $D^{21}\equiv D^>$ are the thermal
Wightman functions and $D^{22}$ is the anti-causal propagator.
\section{Interacting fields in equilibrium states}
In the previous section we have shown, how a technically simple description
in terms of retarded and advanced propagators may be translated into a
causal description through the use of a thermal Bogoliubov
transformation. In the present  section we extend this formalism to 
systems of interacting particles. For simplicity we assume, that other
fields have been integrated out, i.e., the interaction appears in the
form of a non-local self interaction. In general, also
polynomial self-interactions like e.g. a $\phi^4$-theory fall under
our description. 

The interaction is given by an arbitrary interaction Hamiltonian,
which we express in terms of a field operator in the interaction
picture labeled $\widehat{\phi}_x$. In this Hamiltonian,
the field operators are replaced by field amplitudes
belonging to the two representations of the doubled Hilbert space:
\begin{equation}
{\cal H}_{\mbox{\tiny int}}[\phi^\prime_1,\phi^\prime_2] = 
H_{\mbox{\tiny int}}[\widehat{\phi}\rightarrow\phi^\prime_1] - 
H_{\mbox{\tiny int}}[\widehat{\phi}\rightarrow\phi^\prime_2]
\;.\end{equation}
The primes in this equation indicate the fact, that so far we have not
stated the temporal boundary conditions for the fields entering here.
These follow from the almost trivial fact that also the interacting
Green functions have to obey the KMS condition, i.e., the primed
fields must be replaced by differential operators, according to
\begin{equation}
\phi^\prime_i(x) \rightarrow \frac{-{\mathrm i}
  \delta}{\delta j_i(x)} Z_0[j_1,j_2]
\;.\end{equation}
Consequently, the generating functional for Green functions 
in a thermal equilibrium state and containing the proper boundary
conditions is given by
\begin{equation}
\label{ing}
Z\left[j_1,j_2\right] =
\exp\left( - {\mathrm i}\int\!d^4x
 {\cal H}_{\mbox{\tiny int}}\left[
 \frac{ - {\mathrm i}\delta}{\delta j_1},
 \frac{ - {\mathrm i}\delta}{\delta j_2}\right]\right)
  Z_0\left[j_1,j_2\right] 
\;,\end{equation}
where we have to use the final expression for $Z_0$ of eq. (\ref{freeg}).
This generating functional is well known, and has been used very often
in the existing literature. It gives the correct Schwinger-Keldysh
structure of the perturbation series.

However, the virtue of the path integral representation is that it
goes beyond perturbation theory. Thus, we think it useful to 
apply the diagonalization concept of the previous section also to the
interacting case presented here. Naturally this task is much more
tedious than for the case of free fields. We therefore restrict the
discussion to the level which is most important for any
application, i.e., to the level of the full two-point Green function
of the system.

To this end we first define the
generating functional for irreducible vertex functions in the usual
way, cf. ref. \cite[p.476]{IZ80}:
\begin{equation}
\Gamma\left[\phi_1,\phi_2\right]
= \log\left[Z\left[j_1,j_2\right]\right]
 - {\mathrm i}\int\!d^4x\,\left(
 \vphantom{\int}j_1(x)\phi_1(x)- j_2(x)\phi_2(x)\right)
\;.\end{equation} 
$\Gamma$ is the effective action functional for the interacting
fields; for these we introduce a doublet to simplify the
notation:
\begin{equation}
\Phi_x = \left( {\array{l} \phi_1(x) \\ \phi_2(x) \endarray}\right)
\;.\end{equation}
A power series expansion for the effective action functional then yields
\begin{eqnarray}
\nonumber
\Gamma  &=& \frac{1}{2}\partial_t \Phi_x^T\,\tau_3\,\partial_t\,\Phi_x
           -\frac{1}{2}\Phi_x^T\,\tau_3\,\left( -\Delta +
m^2\right)\,\Phi_x \\
           &&-\frac{1}{2}\Phi_x^T\,\int\!\!d^4y\,\Pi(x-y)\,\Phi_y
           + {\cal O}(\Phi^3)
\;.\end{eqnarray}
Here, $\Pi(x-y)$ is the $2\times 2$ matrix valued
one-particle irreducible self energy function
of the system, and $\tau_3 = \mbox{diag}(1,-1)$.
In principle one may also obtain the higher order terms
of such an expansion, but for the purpose of the present paper we
restrict ourselves to the first nontrivial piece involving the self
energy function $\Pi$. 

Since we deal with an equilibrium system, we may assume the self
energy function to depend only on the difference of the space-time arguments.
Consequently, one may perform the Fourier transform into
\begin{equation}
\label{ga2}
\widetilde{\Gamma}(E,\vec{k}) = \frac{1}{2}
\Phi(E,\vec{k})^T\, \left( \tau_3(E^2 - \omega^2_k) 
 -\Pi(E,\vec{k})\right)\,\Phi(E,\vec{k}) 
\;.\end{equation}
In accordance with the previous sections, the doublet
is related to another field $\Xi$ by a 
Bogoliubov transformation \cite{hu92,h94rep}
\begin{eqnarray}
\nonumber
\Phi(E,\vec{k})\hphantom{{}^T\,\tau_3} &=& 
 ({\cal B}(n_B(E)))^{-1}\,\Xi(E,\vec{k})\\
\label{bo}
\Phi(E,\vec{k})^T\,\tau_3 &=& \Xi(E,\vec{k})^T\,\tau_3\,{\cal B}(n_B(E))
\;.\end{eqnarray}
The fields $\Xi$ are therefore those which {\em diagonalize\/} the
effective action functional up to second order.
\begin{eqnarray}
\label{ga3}
&&\widetilde{\Gamma}(E,\vec{k}) = \\
\nonumber
&&\frac{1}{2}\,
\Xi(E,\vec{k})^T\, \tau_3\, \left( 
{\array{lr} E^2 - \omega^2_k -\Pi^R(E,\vec{k}) & \\
& E^2 - \omega^2_k -\Pi^A(E,\vec{k}) \endarray}\right)
\,\Xi(E,\vec{k}) 
\;.\end{eqnarray}
$\Pi^{R}$ and $\Pi^A$  are  the retarded and advanced self energy functions in
momentum space, they are mutually complex conjugate 
and analytical functions of the energy parameter in at least one half plane:
\begin{eqnarray}
\nonumber
\Pi^{R,A}(E,\vec{k}) &=& \mbox{Re}\Pi(E,\vec{k}) \mp {\mathrm i} \pi
  \sigma(E,\vec{k})\\
&=&\int\!\!dE^\prime\,\frac{\sigma(E^\prime,\vec{k})}{
  E-E^\prime \pm {\mathrm i}\epsilon}
\;.\end{eqnarray}
They are also related to the components of the {\em matrix-valued\/}
self energy function by the simple relations
\begin{eqnarray}
\nonumber
\Pi^{R}(E,\vec{k}) &=& 
\Pi^{11}(E,\vec{k}) + \Pi^{12}(E,\vec{k})\\
\label{matrel}
\Pi^{A}(E,\vec{k}) &=& 
\Pi^{11}(E,\vec{k}) + \Pi^{21}(E,\vec{k}) 
\;.\end{eqnarray}
More details of the diagonalization procedure, which (at least to the
level of two-point functions, as considered here) may be formulated
also for non-equilibrium states, can be found in
refs. \cite{hu92,h94rep}.

We proceed by reversing the Legendre transform, i.e., we want to
establish how the interaction affects the diagonalization of the
generating functional for Green functions. This reversal is
straightforward, and one obtains for the generating functional in
terms of the decoupled currents $J_1$ and $J_2$ 
\begin{eqnarray} \nonumber
&&Z\left[J_1,J_2\right] = \\
\nonumber
&&\exp\left[-\frac{{\mathrm i}}{2}
\int\!\!d^4 x\,d^4 y\, \Big(J_{1,x} D^R(x-y) J_{1,y} - 
                       J_{2,x} D^A(x-y) J_{2,y}\Big) \right.\\
\label{diafl}
&&\left.\;\;\;\;\;\;\vphantom{\int}  + {\cal O}(J^3)\right] 
\;.\end{eqnarray}
The retarded and advanced propagator occurring here are, of course,
not the free propagators. Instead they are obtained from a spectral
function ${\cal A}(E,\vec{k})$
\begin{eqnarray}
\nonumber
D^{R,A}(E,\vec{k}) &=& \mbox{Re}D(E,\vec{k}) \mp {\mathrm i} \pi
  {\cal A}(E,\vec{k})\\
\label{self}
&=&\int\!\!dE^\prime\,\frac{{\cal A}(E^\prime,\vec{k})}{
  E-E^\prime \pm {\mathrm i}\epsilon}
\;\end{eqnarray}
which in turn is related to the self energy function components by
\begin{equation}\label{rsc}
{\cal A}(E,\vec{k}) =
   \frac{\sigma(E,\vec{k})}{
         \left(E^2-\omega_k^2-\mbox{Re}\Pi(E,\vec{k}) \right)^2+
         \pi^2\sigma^2(E,\vec{k})}
\;.\end{equation}
The final step of the diagonalization procedure for the interacting
case then is to perform the Bogoliubov transformation to the physical
source currents conforming to the KMS condition. Quite similar to eq.
(\ref{freeg}) we obtain
\begin{equation}
\label{intg}
Z\left[j_1,j_2\right] = 
 \exp\left[-\frac{{\mathrm i}}{2}
\int\!\!d^4 x\,d^4 y\, j_{a,x} D^{ab}(x-y) j_{b,y}
+{\cal O}(j^3)\right] 
\;,\end{equation}
but the matrix valued propagator here is (in momentum space)
\begin{eqnarray}\label{dbk2}
D^{(ab)}(E,\vec{k}) && = \int\limits_{-\infty}^\infty\!\!dE^\prime\,
 {\cal A}(E^\prime,\vec{k})\;\times \\
\nonumber
  &&({\cal B}(n_B(E^\prime)))^{-1}
  \left(\!{\array{ll}
         {\displaystyle \frac{1}{E-E^\prime+{\mathrm i}\epsilon}} & \\
    &    {\displaystyle \frac{1}{E-E^\prime-{\mathrm i}\epsilon}}
\endarray} \right)\;
         {\cal B}(n_B(E^\prime))\,\tau_3
\;.\end{eqnarray}
Consequently, the thermal Bogoliubov transformation (\ref{therb})
diagonalizes the effective action functional up to the level of the
full two-point function also for the interacting case. One may easily
check, that in the limit of a free spectral function the full matrix
valued propagator (\ref{dbk2}) reduces to the free one, eq. (\ref{fg0}).
For the generating functional of Green functions, this diagonalization
implies a factorization into upper and lower contour, see fig. 1.
\section{Generating functional with initial correlations}
The next question to be addressed is how the above formulation has to be
modified in the non-equilibrium case, i.e., how to treat particle
propagation on a non-equilibrium background. To answer it we recall
the principal feature of a non-equilibrium system: The occupation
numbers of single particle states are not fixed a priory. 
For a perturbative treatment this implies some difficulty, since one
may not be able to find operators which ``annihilate'' the
non-equilibrium state.

Consequently one may not use the ``standard'' Wick theorem
to obtain the perturbative propagators. Necessary is a more advanced
formulation of this important tool of perturbation theory. It has
been shown some time ago that such a re-formulation amounts to the
inclusion of {\em initial correlations\/} \cite{FuHa,D84,h90,FW95},
which basically arise when a non-equilibrium ``density matrix'' is
not a Gaussian in the full fields.

The initial correlations appear in a perturbative expansion as
additional vertex functions, coupling one, three, four, five, six
or more fields, but not two fields. 
To explain this in more detail, we leave the path integral formulation
for the moment, and resort to a Fock space picture with free
field operators. To distinguish them from the C-number valued field
amplitudes of the path integral formulation, we denote Fock space
operators by a hat symbol, i.e., $\widehat{\phi}_x$. The expectation
value $\left\langle\cdot\right\rangle$ denotes the trace over this
Fock space with some non-equilibrium density matrix.

In this picture we then apply Wicks theorem to a time-ordered
product of $L$ field operators; resulting in a sum of products of
normal ordered pieces and C-number valued contractions. 
By normal ordering we will henceforth understand to shift all free-field
annihilation operators to the rightmost side, i.e., a normal ordering
with respect to the vacuum. However, our results also hold
for more general normal ordering prescriptions with respect to an
arbitrary equilibrium state:
The argumentation is based on the undebatable fact that 
normal ordering is not possible with respect to a state that has
unknown occupation numbers.

Consequently, in a non-equilibrium state the expectation value of 
a time-ordered product of $L$ free field operators is
\begin{eqnarray}
\nonumber
\left\langle \mbox{T}[\widehat{\phi}^L] \right\rangle 
&=& \sum_{\mbox{\tiny perm}}\sum\limits_{n=0}^{[L/2]}\,
C^{\langle L - 2 n \rangle}\,({\mathrm i}\,D^{11}_{\mbox{\tiny ne}0})^n\\
\label{tp}
&=& \sum\limits_{n=0}^{[L/2]}\,
 \frac{L!}{(L-2n)!\,n!\,2^n}\, 
C^{\langle L - 2 n \rangle}\,({\mathrm i}\,D^{11}_{\mbox{\tiny ne}0})^n
\;,\end{eqnarray}
where $\sum_{\mbox{\tiny perm}}$ 
indicates the sum over all possible permutations of the $L$
space-time coordinates, giving rise to the numerical factor in 
case of symmetrized space-time arguments (second line of
eq. (\ref{tp})).

$D^{11}_{\mbox{\tiny ne}0}$ is the free causal
propagator for the non-equilibrium state and the functions 
$C^{\langle n \rangle}$ are
called correlation kernels \cite{h90}. They may be expressed as a
sum over expectation values of ``vacuum-normal ordered'' products.

The advantage of the correlation kernels is that each of them 
appears only once in a specific order of the perturbative expansion.
Note, that the only correlation kernel not present in  
a general non-equilibrium state is the two-point correlation: It is
absorbed in the propagator of the perturbative expansion. The {\em odd\/}
correlation kernels only appear when the expectation values of the field 
unders consideration are nonzero. In this respect, the inclusion of
initial correlations is a convenient way to obtain a perturbative
expansion also for such fields {\em without\/} previous subtraction of
the expectation value. 

$D^{11}_{\mbox{\tiny ne}0}$ is the 11-matrix element of
\begin{equation}
\label{fcp}
D^{(ab)}_{\mbox{\tiny ne}0}(x,y)
= - {\mathrm i}\, 
\left( { \array{lr}
\left\langle\mbox{T}\left[\widehat{\phi}_x,
  \widehat{\phi}_y\right]\right\rangle & 
  \left\langle\widehat{\phi}_y\widehat{\phi}_x\right\rangle \\
  \left\langle\widehat{\phi}_x\widehat{\phi}_y\right\rangle &
\left\langle\mbox{A}\left[\widehat{\phi}_x,
  \widehat{\phi}_y\right]\right\rangle \endarray } \right) 
\;.\end{equation}
$A[\cdot]$ denotes the anti time-ordered product.
In equilibrium states this reduces to the matrix of eq. (\ref{fg0}).

It is obvious, that the other matrix elements of this propagator occur
in the Wick expansion of totally anti-time ordered, or mixed products
(see ref. \cite{h90} for a complete discussion). It is therefore
important to realize that
the initial correlations do not distinguish between
the two types of indices, i.e., they are {\em not\/} 
contour-ordered functions in the Schwinger-Keldysh formalism.

We now drop the advantage of having a simple perturbative expansion,
and replace the correlation kernels by their expansion into 
expectation values of simple field operator products. It has been
shown in ref. \cite{h90}, that this is also possible in terms of the
Wightman functions rather than in terms of normal ordered products.
This has the advantage that our result becomes independent of the type
of normal ordering prescription we chose.
Indeed, using for the unordered (Wightman) functions the abbreviation
\begin{equation}
P^{\langle L\rangle}_{1\dots L}={\langle\widehat{\phi}_1\dots
\widehat{\phi}_L \rangle}
\;\end{equation}
we may write
\begin{equation}
\label{corr}
C^{\langle L  \rangle}=\sum\limits_{n=0}^{[L/2]}\,
\frac{L !}{(L-2n)!\,n!\,(-2)^n}\,
P^{\langle L - 2 n \rangle}\,
  \left\langle\widehat{\phi}\widehat{\phi}\right\rangle^n
\;.\end{equation}
Again, the numerical factor is due to the sum over all permutations of
the space-time arguments, i.e., we understand the above expression to
be symmetrized in these.

The reason for this rewriting is, that one may easily obtain 
the generating functional for the correlation kernels in terms of
the initial state Wightman functions,
\begin{equation}
\label{gfc}
W\left[\eta\right] =
\exp\left(+\frac{1}{2}\int\!\!d^4x d^4y\;\eta_x \left\langle
 \widehat{\phi}_x\widehat{\phi}_y\right\rangle \eta_y \right)\,
\left\langle \exp\left({\mathrm i}\int\!\!d^4z\,\eta_z\,
  \widehat{\phi}_z\right)\right\rangle
\,.\end{equation}
The correlation kernels are obtained from this 
by subsequent derivatives, i.e.,
\begin{equation}
C^{\langle K\rangle}_{1\dots K} =
\left(\frac{ (-{\mathrm i})^K\delta^K}{\delta \eta_1 \dots \delta \eta_K}\,
W\left[\eta\right]\right)_{\eta\equiv 0}
\,,\end{equation}
The generating functional of the time-ordered functions therefore is
\begin{eqnarray}
Z^{11}[\eta] &=& \left\langle \mbox{T}\left[\exp \left(
 {\mathrm i}\int\!\! d^4z\; \eta_z
\widehat{\phi}_z\right)\right] \right\rangle\nonumber \\
&=& 
\exp\left(-\frac{1}{2}\int\!\!d^4x d^4y\;\eta_x\left
({\mathrm i}D^{11}_{\mbox{\tiny ne}0}(x,y)- \left\langle
\widehat{\phi}_x\widehat{\phi}_y\right\rangle\right)\eta_y\right)\nonumber\\
&&\;\;\;\;\;\;\times\left\langle \exp\left({\mathrm i}\int\!\!d^4z\,\eta_z\,
  \widehat{\phi}_z\right)\right\rangle \nonumber\\
\label{znc}
&=&
\exp\left[-\frac{1}{2}\int\!\!d^4x d^4y\;\eta_x\, 
 {\mathrm i}D^{11}_{\mbox{\tiny ne}0}(x,y)\, \eta_y\right]\times W[\eta]
\;.\end{eqnarray}
From these expressions it is obvious how the equilibrium nature of a
state is connected to the vanishing of the correlation kernels $C$:
It is a gaussian of the fields $\phi$, and therefore the generating
functional $W$ is identical to 1.

Finally, we extend this formula to the matrix valued formulation. As 
we have indicated above, this brings into play the other matrix
elements of eq. (\ref{fcp}). Furthermore the correlation kernels are
identical for both types of indices, thus one
has to replace the current $\eta$ in the functional $W[\eta]$ by the sum of
$j_1$ and $j_2$. The interaction is introduced in a perturbative way
as in eq. (\ref{ing}). The generating functional for non-equilibrium Green
functions therefore is
\begin{eqnarray}
\nonumber
Z_{\mbox{\tiny ne}}\left[j_1,j_2\right]& =&
\exp\left( - {\mathrm i}\int\!\!d^4u\;
 {\cal H}_{\mbox{\tiny int}}\left[
 \frac{ - {\mathrm i}\delta}{\delta j_{1,u}},
 \frac{ - {\mathrm i}\delta}{\delta j_{2,u}}\right]\right)\times\\
\nonumber
&&
 \exp\left[-\frac{1}{2}
\int\!\!d^4 x\,d^4 y\; j_{a,x}\left( {\mathrm i}D^{ab}_{\mbox{\tiny
ne}0}(x,y)  
  -\left\langle \widehat{\phi}_x\widehat{\phi}_y\right\rangle\right)
\,j_{b,y}\right]\times\\ 
&&
 \label{negf}
 \left\langle
 \exp\left[{\mathrm i}
\int\!\!d^4 z\; (j_{1,z} + j_{2,z})\,\widehat{\phi}_z\right]\right\rangle
\;.\end{eqnarray}
This allows to recover the full perturbation expansion
for non-equilibrium Green functions in an arbitrary state. 
States with nonzero field expectation value as well as states with
complicated correlations are handled correctly by this expression.

However, missing in this framework is the connection to the preceding
sections, i.e., to the question of {\em temporal\/} boundary
conditions. This gap will be closed in the following section.
\section{Cumulant expansion and path integrals}
The expression for generating functional of free 
non-equilibrium Green functions in an arbitrary system is given 
in Eq. (\ref{skgf}). We now proceed
along the line of thought underlying the {\em path integral\/}
formulation: Fields are expanded around their classical value, i.e.,
we set
$\phi_x=\phi_x^{\mbox{\tiny cl}}+\phi_x^{\mbox{\tiny q}}$.
The classical field $\phi^{\mbox{\tiny cl}}$ is the solution of
the field equation with an external current as source 
\begin{equation}
\left( \Box_x + m^2 \right)\phi_x^{\mbox{\tiny cl}} =  J_{x} \, .
\end{equation}
Inserting this equation into 
(\ref{skgf}) allows to rewrite the generating functional as
\begin{eqnarray}
Z_0\left[J \right] &=& \int\!\!{\cal D}[\phi^{\mbox{\tiny q}}]\,
 \exp\left[\frac{{\mathrm i}}{2}\int\limits_{\cal C}\!\!d^4 x \left( 
 \phi_x^{\mbox{\tiny q}}
  \left( \Box_x + m^2 \right)\phi_x^{\mbox{\tiny q}} 
 \right)\right] \nonumber\\
\label{zcl}
& &\;\;\;\;\;\times 
 \exp\left[\frac{{\mathrm i}}{2}\int\limits_{\cal C}
 \!\!d^4 x \left( \phi_{x}^{\mbox{\tiny cl}} J_{x}
\right)\right]  
\,.\end{eqnarray}
The first factor is a trivial Gaussian integral, i.e., it amounts to a
phase of the whole expression and will be set to 1 henceforth. For
$\phi^{\mbox{\tiny cl}}$ in the second factor the classical solution 
obeying the correct boundary conditions must be substituted. In the case of
equilibrium systems we may split up the fields and currents due to
their location on the contour ${\cal C}$ depicted in fig. 1:
\begin{equation}
\label{pcl}
\phi_x^{\mbox{\tiny cl}} = - \int\!\!d^4 y\; 
\left\{ {\array{ll} D^R(x-y) J_{1,y} & \mbox{ if $x$ on upper contour } \\
                    D^A(x-y) J_{2,y} & \mbox{ if $x$ on lower contour.} 
\endarray } \right.
\end{equation}
This reproduces the generating functional from eq. (\ref{fjgf}).
Another possibility for this splitting is the causal boundary condition
\begin{equation}
\label{pcl2}
\phi_x^{\mbox{\tiny cl}} = - \int\!\!d^4 y\; 
\left\{ {\array{ll} D^{1a}(x-y) j_{a,y} & \mbox{ if $x$ on upper contour } \\
                    D^{2a}(x-y) j_{a,y} & \mbox{ if $x$ on lower contour,} 
\endarray } \right.
\end{equation}
which results in the expression (\ref{freeg}) for the generating
functional.

Here therefore is the place, where the temporal boundary condition
enters the calculation of $Z[j]$. We now ask the question, how this 
classical equation has to be modified for non-equilibrium states.

To this end, we introduce the irreducible {\em cumulants\/} by
relating them to the generating functional of the correlation kernels
\cite{FW96}:
\begin{eqnarray}
\nonumber
W\left[\eta\right] &=& {\mathrm i}\int\!\!d^4x\,
  \left\langle\widehat{\phi}_x\right\rangle\eta_x
+\sum\limits_{n=3}^\infty\,\int\!\!d^4 1\dots d^4 n\,
 \frac{{\mathrm i}^n}{n !}\,
  C^{\langle n\rangle}\, \eta_1\dots \eta_n\\
&=&\exp\left[{\mathrm i}\int\!\!d^4x\,
  \left\langle\widehat{\phi}_x\right\rangle\eta_x
+\sum\limits_{n=3}^\infty\,\int\!\!d^4 1\dots d^4 n\,
  \frac{{\mathrm i}^n}{n !}\,
  \kappa(1,\dots n)\, \eta_1\dots \eta_n \right]
\,.\end{eqnarray}
In other words, the logarithm of the generating functional for
cumulants equals the generating functional for the correlation
kernels.

These cumulants $\kappa(1,\dots n)$ appear more than once in a given order of 
perturbation theory, whereas the correlation kernels appear only once.
The use of cumulants therefore allows a 
resummation of the perturbation series similar to the Dyson equation;
as well as the simple formulation of a linked cluster theorem \cite{FW96}.

For completeness we quote the expectation value of a time-ordered
product in terms of these cumulants, i.e., the equivalent of eq.
(\ref{tp}):
\begin{equation}
\left\langle \mbox{T}[\widehat{\phi}^L] \right\rangle 
= \sum_{\mbox{\tiny perm}}\sum\limits_{n=0}^{[L/2]}\,\sum\limits_{j=0}^{L-2n}
\,\kappa\{ \widehat{\phi}^{L-2n}\}_j
\,\left({\mathrm i}\,D^{11}_{\mbox{\tiny ne}0}- 
  \left\langle\widehat{\phi}\right\rangle
  \left\langle\widehat{\phi}\right\rangle\right)^n
\;,\end{equation}
where $\kappa\{ \widehat{\phi}^{L}\}_j$ denotes a product of $j$
cumulants not containing 2-point cumulants whose arguments are a
partition of the $L$ arguments into $j$ disjoint sets.

Inserting these definitions in eq. (\ref{znc}), we obtain for the
causal generating functional in terms of the cumulants
\begin{eqnarray}
\nonumber
Z_{\mbox{\tiny ne}0}[j_1,j_2]& =&
 \exp\left[-\frac{1}{2}\int\!\!d^4x d^4y\;\,j^a_x\, 
 {\mathrm i}D^{ab}_{\mbox{\tiny ne}0}(x,y)\, j^b_y \right. \\
\nonumber
&+&{\mathrm i}\int\!\!d^4x \,
  \left\langle\widehat{\phi}_x\right\rangle(j_{1,x}+j_{2,x})\\
\label{zncc}
&+&\left.\sum\limits_{n=3}^\infty\,\int\!\!d^4 1\dots d^4 n\,
  \frac{{\mathrm i}^n}{n !}\,
  \kappa(1,\dots n)\, \prod\limits_i^n(j_{1,i}+j_{1,i}) \right]
\;,\end{eqnarray}
where $j_{1,i}\equiv j_1(x_i), i=1\dots n$.
Note, that one may subtract the cumulant 
$\kappa(x,y)=\left\langle\widehat{\phi}_x\widehat{\phi}_y\right\rangle$
from the
propagator and then obtains a complete series over all values of $n$
for the last term. Also, the introduction of an interaction may be
carried out as presented in section 3.

We now use the results of section 2 to make the
transformation into the retarded/advanced picture involving the
currents $J_x$. Although in section 2 we deal with an equilibrium
state, it is clear that there exists a similar diagonalization
transformation for the non-equilibrium case. To see this more clearly,
we perform the Fourier transform of eq. (\ref{fcp}) with respect to
$(x-y)\rightarrow P$, and write it in terms of the two arguments $P$
and $X=(x+y)/2$. It follows that
\begin{eqnarray}
\nonumber
&&{\cal B}(N_{XP}) \,\left(\vphantom{\int}D^{(ab)}\right)\,
 \tau_3\,\left({\cal B}(N_{XP})\right)^{-1} 
=\\
&& \left( {\array{lr}
    D^R_{XP} & \;\;\left(N_{XP} D^{21}_{XP} - (1+N_{XP}) D^{12}_{XP}\right) \\
    0        & D^A_{XP} \endarray} \right)
\;,\end{eqnarray}
where we may chose $N_{XP}$ in such a way that also the upper right
corner of the propagator matrix is zero. In ref. \cite{h94rep} is is
shown, that the corresponding $N_{XP}$ fulfills a transport equation
and in the equilibrium limit approaches a Bose-Einstein distribution
function. It is therefore indeed the generalized non-equilibrium
distribution function.

However, at the level of the external sources $j$, $J$ this
transformation has the disadvantage to be more complicated than
eq. (\ref{therb}). It mixes various momentum components of the 
$j_i$ to obtain the $J_i$ and vice versa \cite{h94rep}:
\begin{eqnarray}
&&\int\!\!d^4 x d^4 y\;j_{a,x}\,D^{(ab)}(x,y)\,j_{b,y}\\
\nonumber
&&\;\;=\int\!\!\frac{d^4 q}{(2\pi)^4} \frac{d^4 p}{(2 \pi)^4} d^4 X
  \;{\mathrm e}^{-{\mathrm i}(q-p)X}\,
  \underbrace{
  \left({\array{l} j_{1,q} \\ j_{2,q}\endarray}\right)^T\!\!\!
  \,\tau_3\,\big({\cal B}(N(X,(p+q)/2))\big)^{-1}\,\tau_3}\\  
&&\nonumber
\times
  \left( {\array{lr}
    D^R(X,(p+q)/2)\!\!\!\!\! & 0\\
    0        & D^A(X,(p+q)/2) \endarray} \right)
  \underbrace{{\cal B}(N(X,(p+q)/2))
  \left({\array{l} j_{1,q} \\ j_{2,q}\endarray}\right)}
\;.\end{eqnarray}
The ``new'' currents are marked by the horizontal braces. Let us note
at this point, that  the momentum mixing Bogoliubov transformation
naturally couples the system to gradients in the generalized
distribution function. This allows to calculate transport coefficients
\cite{h94rep} as well as non-equilibrium effects in 
two-particle interferometry \cite{hhbr95}. 

This separation however fails when taking into account the higher
order terms, i.e., those with correlations built in.
Consequently, for a general non-equilibrium state the generating
functional for Green functions does not factorize into a part of the
upper branch and one on the lower branch of fig.1.
This may also be seen from the equation that is fulfilled by 
the classical field $\phi^{\mbox{\tiny cl}}$:
\begin{eqnarray}
\nonumber
\phi_x^{\mbox{\tiny cl}} &=& 
 -{\mathrm i}\frac{\delta}{\delta j_{1,x}}\,Z[j_1,j_2] \\
\nonumber
&=&\sum_{n=1}^\infty \frac{{\mathrm
  i}^{n-1}}{(n-1)!} \int\limits_{\cal C}
\!\!d^4 1\dots d^4 n\; J_1\dots J_n 
 \;\left\langle \mbox{T}
 \left[\widehat{\phi}_1 \dots
\widehat{\phi}_n \widehat{\phi}_x \right]\right\rangle\\
\nonumber 
 &=& \left\langle \widehat{\phi}_x\right\rangle - 
\int\!\!d^4y \;  D^{1a}(x,y) j_{a,y}\\
& &\qquad+\sum_{n=3}^\infty \frac{{\mathrm i}^n}{n!} \int
\!\!d^4 1\dots d^4 n \; \prod\limits_i^n(j_{1,i}+j_{2,i})\kappa(1,\dots,n,x)
\;.
\end{eqnarray}
This equation contains the boundary condition for the {\em
classical\/} field $\phi^{\mbox{\tiny cl}}$. Its is also obtained
by application of a linked cluster theorem to the perturbation
expansion, i.e.,
when evaluating the perturbation series for the classical field amplitude
by hand considering only contributions from connected diagrams.

Furthermore, it is obvious that it also approaches the
Kubo-Martin-Schwinger (KMS) boundary condition (periodicity in the
imaginary time direction) in the limit of zero
cumulants, because it is then equivalent to eq. (\ref{pcl2}). 
\section{Conclusion}
With the present paper we achieved two goals. First of all we have
extended the concept of {\em diagonalized\/} matrix valued propagators
into the path integral formulation of quantum field theory. We were
able to show, how to construct a simple perturbative expansion involving only
retarded and advanced Green functions. This diagonalization procedure
exploits a thermal Bogoliubov symmetry of the system \cite{hu92}.
The presence of this symmetry allows even simpler thermal Feynman
rules when switching to a lower dimensional representation of the 
symplectic group \cite{h94rep}. 

The diagonalization of the action integral amounts to a factorization
of the generating functional for Green functions. 
Our result therefore includes the one of ref. \cite{EVANS}, but is
more general since we have achieved the factorization also for the
full two-point function.   

The second major goal we have achieved with the present paper is the inclusion
of initial correlations into the path integral framework. This might
be a future asset for two very different reasons. On one hand it
allows to obtain a perturbative expansion also for truly 
non-equilibrium systems, where the boundary conditions for particle
propagation are fuzzy at best. To this end one may use the results of
section 4, where the correlations were introduced via the {\em initial
state Wightman functions\/}. In particular, one may think of a
practical application to transport theory, where the perturbative
treatment only works for some time. Thereafter one has to requantize
the system with new degrees of freedom, e.g., with new
quasi-particles. For the next time step then the correlations of the
initial state play the role of additional vertices \cite{h90}.

On the other hand, the correlations were related to a cumulant
expansion of the action integral \cite{FW96}. In this formulation, as
presented in the last section of this paper, we were able to write
down the integral equation fulfilled by the classical field amplitude.
It is this integral equation which proves the close relationship
between initial correlations and the boundary condition in time that
one has to implement in quantum field theory for non-equilibrium
states.

We have furthermore proven, that the path integral for quantum fields
in non-equilibrium states does not factorize into pieces on the upper
and lower Keldysh contour.

\setlength{\unitlength}{1mm}
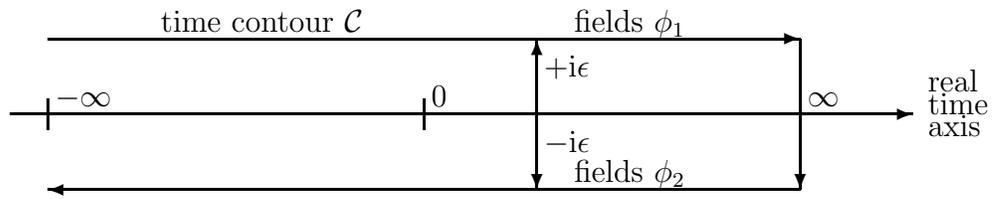
\begin{figure}
\begin{picture}(140,60)
\thicklines
\put(10,30){\vector(1,0){120}}
\put(132,33){real}
\put(132,30){time}
\put(132,27){axis}
\put(15,32){\line(0,-1){4}}
\put(16,31){$-\infty$}
\put(65,32){\line(0,-1){4}}
\put(66,31){0}
\put(116,31){$\infty$}
\put(15,40){\vector(1,0){100}}
\put(115,40){\vector(0,-1){20}}
\put(115,20){\vector(-1,0){100}}
\put(30,41){time contour ${\cal C}$}
\put(85,41){fields $\phi_1$}
\put(85,21){fields $\phi_2$}
\put(80,30){\vector(0,1){10}}
\put(81,35){$+{\mathrm i}\epsilon$}
\put(80,30){\vector(0,-1){10}}
\put(81,25){$-{\mathrm i}\epsilon$}
\end{picture}
\caption{Time path for the action integral entering the generating functional.}
\end{figure}
\end{document}